\documentclass[twocolumn,showpacs,preprintnumbers]{revtex4}

\usepackage{graphicx}
\usepackage{bm}

\begin{document}

\title{Magnetic field induced confinement-deconfinement transition in
graphene quantum dots}

\author{G. Giavaras}
\affiliation{Department of Physics and Astronomy, University of Leicester, 
Leicester LE1 7RH, UK}

\author{P. A. Maksym}
\affiliation{Department of Physics and Astronomy, University of Leicester, 
Leicester LE1 7RH, UK}

\author{M. Roy}
\affiliation{Department of Physics and Astronomy, University of Leicester, 
Leicester LE1 7RH, UK}

\date{\today}

\begin{abstract}
Massless Dirac particles cannot be confined by an electrostatic potential.
This is a problem for making graphene quantum dots but confinement can be
achieved with a magnetic field and here, general conditions for confined
and deconfined states are derived. There is a class of potentials for which
the character of the state can be controlled at will. Then a
confinement-deconfinement transition occurs which allows the Klein paradox
to be probed experimentally in graphene dots.
\end{abstract}


\maketitle

Single layer graphene is attracting attention because its charge carriers
are massless, relativistic particles \cite{Geim07}. The relativistic
effects result from a unique, zero-gap band structure that leads to quantum
states described by the two-component Dirac-Weyl equation. This allows
relativistic physics to be explored in a solid state system and has many
potential applications ranging from high frequency electronics
\cite{Geim07} to quantum computing \cite{Trauzettel07}. In particular,
graphene quantum dots are very attractive as spin qubits because they are
expected to have low spin decoherence \cite{Trauzettel07}. However there is
a problem with making graphene dots for quantum computing or any other
application: relativistic effects prevent massless particles from being
confined by an external potential. 

The problem results from the Klein paradox
\cite{Katsnelson06,Peres06,Cheianov06}. When relativistic particles
with mass are incident on a 1D potential barrier, the state in the barrier
decays exponentially unless the barrier height exceeds the threshold for
pair production, at which point the state in the barrier becomes oscillatory.
The paradox is that any attempt to enhance the localisation by increasing
the barrier height eventually destroys it. But there is no threshold for
pair production for massless particles so exponential decay and bound
states do not occur.

Graphene dots can be formed from external potentials or nanocrystals.
The quantum states \cite{Silvestrov07,Martino07,Chen07,Matulis07},
in external potentials \cite{nanocrystal} are quasi-bound:
they have a low amplitude oscillatory tail and are similar to the
scattering resonances studied in undergraduate physics. A perpendicular
magnetic field enhances the localisation of these states \cite{Chen07} and
true bound states can occur in graphene dots defined by a spatially
non-uniform field \cite{Martino07}. So a magnetic vector potential has a
localising effect that tends to cancel the delocalising effect of a scalar
potential. But what are the general conditions for confined states to occur
when an electrostatic scalar potential and a magnetic vector potential are
applied to graphene simultaneously?

This question is answered in the present Letter. It is shown that both true
bound states and quasi-bound states occur, depending on the form of the
potentials. In addition, there is a third and most interesting possibility.
In some cases, the character of the states depends on the parameters of the
potentials and can be controlled at will. A confinement-deconfinement
transition then occurs in which the character of the states changes from
oscillatory to exponential as in the Klein paradox for particles with
mass. This gives a way of probing the Klein paradox experimentally in a
solid state system and numerical studies of the quantum states in a
realistic dot model show it is feasible.

The system considered here is a cylindrically symmetric graphene dot. The
dot region is defined by an electrostatic potential, $V(r)$,
and the magnetic vector potential, 
$A_{\theta}(r)$, is in the azimuthal ($\theta$) direction. The functional
form of these potentials has a
significant effect on the character of the states. In particular, when
$V$ and $A_{\theta}$ increase as power laws,
$V = V_0r^s$, $A_{\theta} = A_0r^t$, $s,t >0$, 
the character of the states depends
critically on $s$ and $t$. If $s > t$ the states oscillate in the
asymptotic regime of large $r$ but decay exponentially when $s < t$. In
both cases the asymptotic character of the states is independent of $V_0$
and $A_0$ but when $s=t$ the character of the states does depend on these
constants and a transition from exponential to oscillatory behaviour occurs
when $V_0$ is increased or $A_0$ is decreased. This is the
confinement-deconfinement transition.

The two-component envelope function, $\psi$, satisfies $(V + \gamma
\bm{\sigma}\cdot \bm{\pi}/\hbar)\psi = E\psi$, where $\bm{\sigma}$ are
the Pauli matrices, $\bm{\pi} = \mathbf{p} + e\mathbf{A}$ and $\gamma =
646$ meV nm \cite{Chen07}. For cylindrically symmetric systems $\psi =
(\chi_1(r)\exp(i(m-1)\theta), \chi_2(r)\exp(i m\theta))$ where $m$ is
the angular momentum quantum number. The radial functions $\chi_1$
and $\chi_2$ satisfy
\begin{eqnarray}
\frac{V}{\gamma}\chi_1 - i\frac{d\chi_2}{dr} -i \frac{m}{r}\chi_2 -
i\frac{e}{\hbar}A_{\theta}\chi_2 &=& \frac{E}{\gamma}\chi_1,\label{radeq1}\\ 
- i\frac{d\chi_1}{dr} +i \frac{(m-1)}{r}\chi_1 +
i\frac{e}{\hbar}A_{\theta}\chi_1 + \frac{V}{\gamma}\chi_2 &=& 
\frac{E}{\gamma}\chi_2.\label{radeq2}
\end{eqnarray}
These equations can be uncoupled by differentiating them and this leads
to the relation
\begin{equation}
\chi_2'' + a(r)\chi_2' + b(r)\chi_2 =0,
\label{chi2eq}
\end{equation}
where
\begin{eqnarray}
a(r) &=&\frac{1}{r} + \frac{1}{E-V}\frac{dV}{dr},\label{adef}\\
b(r) &=& -\frac{m^2}{r^2} +
\left(\frac{m}{r} +\frac{e}{\hbar}A_{\theta}\right) \frac{1}{E-V}\frac{dV}{dr}
-\frac{(2m-1)}{r}\frac{e}{\hbar}A_{\theta}\nonumber\label{bdef}\\
&+&\frac{e}{\hbar}\frac{dA_{\theta}}{dr}
+\frac{(E-V)^2}{\gamma^2} - \frac{e^2}{\hbar^2}A_{\theta}^2.
\end{eqnarray}
The first order derivative in Eq.~(\ref{chi2eq}) is eliminated
by putting $\chi_2(r) = u_2(r)\exp(-\int a(r)dr/2)$ which gives
\begin{equation}
u_2'' + k_2^2(r) u_2 = 0,
\label{ueq}
\end{equation}
where $k_2^2(r)= b - a'/2 -a^2/4$. Although $k_2^2$ diverges when $E=V$,
$\chi_2$ is regular there. Since $\exp(-\int a(r)dr/2)$ is not an
oscillatory function of $r$, $u_2$ has the same character as $\chi_2$.
Eq.~(\ref{ueq}) shows that this character is oscillatory when $k_2^2$ is
positive and exponential when $k_2^2$ is negative. Asymptotic exponential
decay is
characteristic of a bound state but here ``confined'' is used
to indicate a bound state that is localised near the centre of the dot.

When $V$ and $A_{\theta}$ increase as
power laws, the asymptotic form of $k_2^2$ is
$(V_0/\gamma)^2r^{2s} - (eA_0/\hbar)^2r^{2t}$.
Hence $k_2^2$ is positive when $s>t$, leading to oscillatory character
and deconfined states. And $k_2^2$ is negative when $s<t$, leading to
exponential character and confined states. However if $s=t$ the
asymptotic form is $[(V_0/\gamma)^2- (eA_0/\hbar)^2]r^{2t}$ so the sign of
$k_2^2$ depends on $V_0$ and $A_0$ and a confinement-deconfinement
transition occurs when $V_0^2= (\gamma eA_0/\hbar)^2$. This also follows
from the uncoupled equation for $\chi_1$: the corresponding
$k^2$-value, $k_1^2 \ne k_2^2$ but $k_1^2\rightarrow k_2^2$ in the
asymptotic limit so $\chi_1$ and $\chi_2$ have the same character.

To investigate the states further, Eqs.~(\ref{radeq1}) and (\ref{radeq2})
are solved numerically. The Hamiltonian, $H$, satisfies
\begin{eqnarray}
&&\int_0^R\int_0^{2\pi} \left[\psi^*_{\alpha} H \psi_{\beta} - 
(H \psi_{\alpha})^* \psi_{\beta}\right] d\theta rdr =\nonumber\\
&&-2\pi i\gamma\left[(\chi_{1\alpha}^*\chi_{2\beta}+
\chi_{2\alpha}^*\chi_{1\beta})r\right]^R_0, \label{hermiteq}
\end{eqnarray}
where $\psi_{\alpha}$ and $\psi_{\beta}$ are two-component states of
angular momentum $m$ and $\chi_{i\alpha}$,$\chi_{i\beta}$ are the
corresponding radial functions. Eqs.~(\ref{radeq1}) and (\ref{radeq2})
lead to a Hermitian eigenvalue problem when the boundary terms in 
Eq.~(\ref{hermiteq}) vanish. For this to happen it is sufficient
that one component is regular at the origin and one component vanishes
at the boundary, $R$. Then it follows from
Eqs.~(\ref{radeq1}) and (\ref{radeq2}) that both components are regular
at the origin but it does not follow that both components vanish at $R$. 
However a true bound state has an exponential tail so both components of
a bound state vanish in the limit of large $R$. Hence
confined and deconfined states can be distinguished 
by solving Eqs.~(\ref{radeq1}) and (\ref{radeq2}) subject to the boundary
conditions that one component is regular at the origin and \textit{one}
component
vanishes at $R$ and then looking for an exponential tail in \textit{both}
components.

\begin{figure}
\begin{center}
\includegraphics[height=11.5cm]{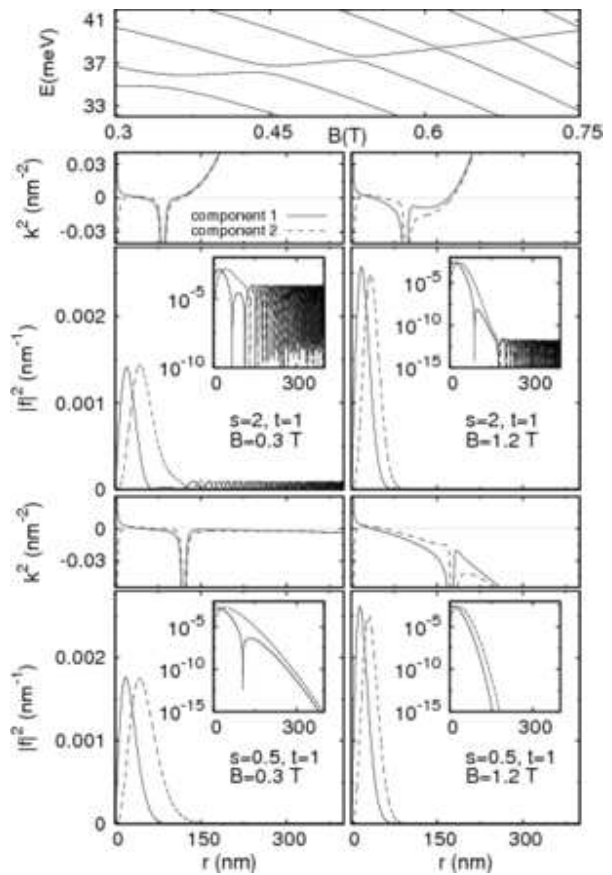}
\caption{Confined and deconfined states for cases when 
no transition occurs. The frame above each state shows $k_i^2$.
Topmost frame: $E(B)$ for $s=2$, $t=1$.}\label{notrans}
\end{center}
\end{figure}

Eqs.~(\ref{radeq1}) and (\ref{radeq2}) are solved by discretizing them
on a uniform grid. By applying the time reversal operator to
Eqs.~(\ref{radeq1}) and (\ref{radeq2})
it can be shown that $E(m,A_0) = E(1-m,-A_0)$. It is important to ensure
that the eigenvalues of the 
discretized Hamiltonian have the same property and this requires
identical numbers of grid points for $\chi_1$ and $\chi_2$.
This excludes the use of centred differences
\cite{reason1} so $d/dr$ is approximated by the forward difference
operator $L_f$ or the backward difference operator $L_b$. The procedure
depends on $m$. When $m\le 0$, $\chi_2(R)$ is chosen to vanish
\cite{reason2}, $L_f$ is used to find $d\chi_2/dr$ and $L_b$ to find
$d\chi_1/dr$ and vice-versa when $m\ge 1$. Although this 
guarantees that $E(m,A_0) = E(1-m,-A_0)$, it has the disadvantage that
numerical errors
are linear in the step length $\Delta r$. To compensate for this, $\Delta
r$ is kept small and all the eigenvalues computed in this work
are accurate to about 1 part in $10^3$, except for $s=2,t=1$ and
$s=2,t=2$, where there
are rapid oscillations but the accuracy is still better than 2\%.
The discretization leads to a non-Hermitian matrix eigenvalue problem. A
similarity transformation is used to reduce this to a real, symmetric
eigenvalue problem which is solved numerically.

The quantum states shown in the present work are all $m=1$ states. Other
states exhibit similar features, although the amplitude of oscillations is
$m$-dependent \cite{Chen07}. Since the main focus of this work is on
the confinement-deconfinement transition, the states are selected so that
the behaviour at the origin does not change significantly when the
potential parameters are changed. All the states have been selected to have
a large amplitude close to the origin and can be regarded as dot states.
The energies as a function of magnetic
field typically show a series of anti-crossings (Fig.~\ref{notrans}) and
where necessary, the line of anti-crossings is followed to preserve the
qualitative form of the state at the origin \cite{anti}. Every energy
computed here is between 26 and 67 meV, within the validity limit of 
the linear graphene Hamiltonian ($\approx \pm 1$ eV \cite{Geim07}).

\begin{figure}
\begin{center}
\includegraphics[height=10.0cm]{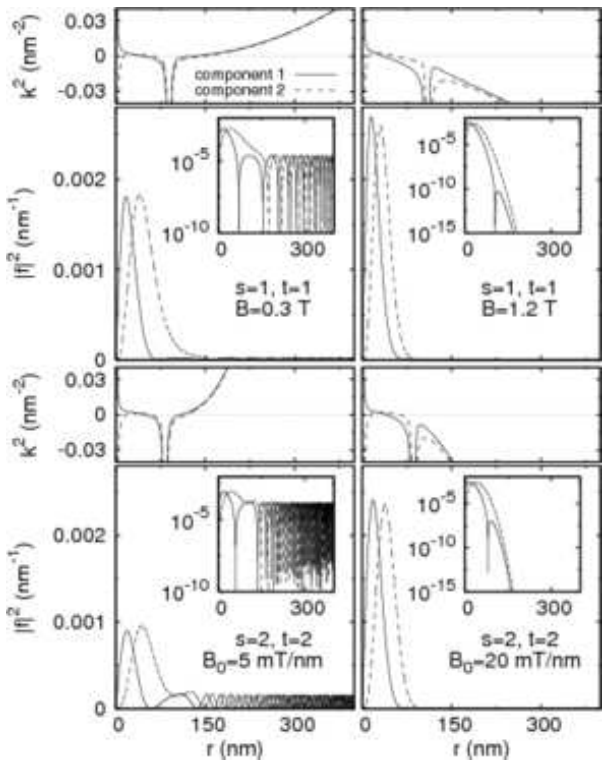}
\caption{As Fig.~\ref{notrans} but for two cases
which exhibit a
confinement-deconfinement transition.}\label{trans}
\end{center}
\end{figure}

Confined and deconfined states are illustrated in Fig.~\ref{notrans}.
The radial probability distribution, $|f_i|^2 \equiv r|\chi_i|^2$,
$i=1,2$ is shown together with $k_i^2$. The insets show $|f_i|^2$ on a
logarithmic scale. $R=600$ nm, large enough to ensure that the asymptotic
sign of $k_i^2$ has been reached. The magnetic field, $B$, is uniform.
When $s=2$, $t=1$,
$V_{0}=5\times10^{-3}$ meV nm$^{-2}$, the asymptotic sign of $k_i^{2}$ is
positive and the asymptotic character of $|f_i|^2$ is oscillatory
independent of $B$.
The amplitude of the oscillations decreases with increasing $B$.
When  $s=0.5$, $t=1$, $V_{0}=5$ meV nm$^{-1/2}$ and $B\ne 0$, the
asymptotic sign of $k_i^2$ is negative and the asymptotic
character of $|f_i|^2$ is exponential, as can be seen from the insets.

The confinement-deconfinement transition is illustrated in
Fig.~\ref{trans}. $R=600$ nm, again large enough to reach the asymptotic
regime. When  $s=1$, $t=1$, $V_{0}=0.5$ meV nm$^{-1}$ and $B=0.3$ T, the
asymptotic sign of $k_i^{2}$ is positive and the asymptotic character
is oscillatory. In contrast, when 
$B=1.2$ T, the asymptotic sign of $k_i^2$ is negative and the asymptotic
character is exponential.
The transition also occurs in non-uniform magnetic fields and it may be
possible to generate a suitable field by putting a dot under a
superconducting obstacle in a uniform field. Fig.~\ref{trans} (bottom)
shows the transition for $s=2$, $t=2$, $V_{0}=5\times10^{-3}$ meV nm$^{-2}$,
that is parabolic confinement in a linearly increasing field, $B(r) = B_0 r$.

\begin{figure}
\begin{center}
\includegraphics[height=3.6cm]{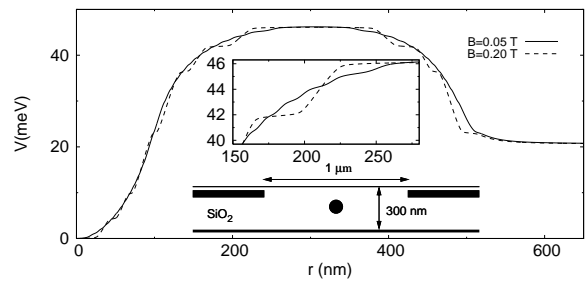}
\caption{$V(r)$ for the gate geometry shown in
the bottom inset.}\label{potent}
\end{center}
\end{figure}
In any real quantum dot, the scalar potential would approach a finite
asymptotic value instead of increasing without limit. Consequently, all
the states of a graphene dot in a magnetic field have an exponential tail.
However, an effect similar to the confinement-deconfinement
transition occurs in the middle distance region between the
centre of the dot and the asymptotic regime.

For this transition to be observable, the dot level has to be in the region
of very low density of states between the bulk Landau levels. This requires
a potential similar to the one shown in Fig.~\ref{potent}. The asymptotic
value of the potential is engineered to be just below the dot level. This
puts the dot level between the zeroth and first Landau levels. Thus the dot
level can be isolated from the bulk Landau levels provided that they are
narrow enough.

The required potential can be generated by gate electrodes. One possible
arrangement is a metal plate with a circular hole that contains an
electrode. The graphene sheet is above these electrodes on 300 nm of
SiO$_2$ on a back-gate at 0 V. The plate (-1 V) generates the
asymptotically flat part of the potential, the hole generates the barrier
and the central electrode (-2 V) generates the well. Similar gated,
monolayer graphene nanostructures have been fabricated recently
\cite{Huard07}.  The potential in Fig.~\ref{potent} was computed by solving
the Poisson equation on a discrete mesh.  Screening by the graphene sheet
was treated in the Thomas-Fermi approximation \cite{DiVincenzo84}. The
resulting potential is magnetic field dependent because the density of
states is field dependent. This causes steps in the potential (inset to
Fig.~\ref{potent}) which occur when the number of occupied Landau levels
changes. However the field dependence is weak at the low fields considered
here.

\begin{figure}
\begin{center}
\includegraphics[height=10.0cm]{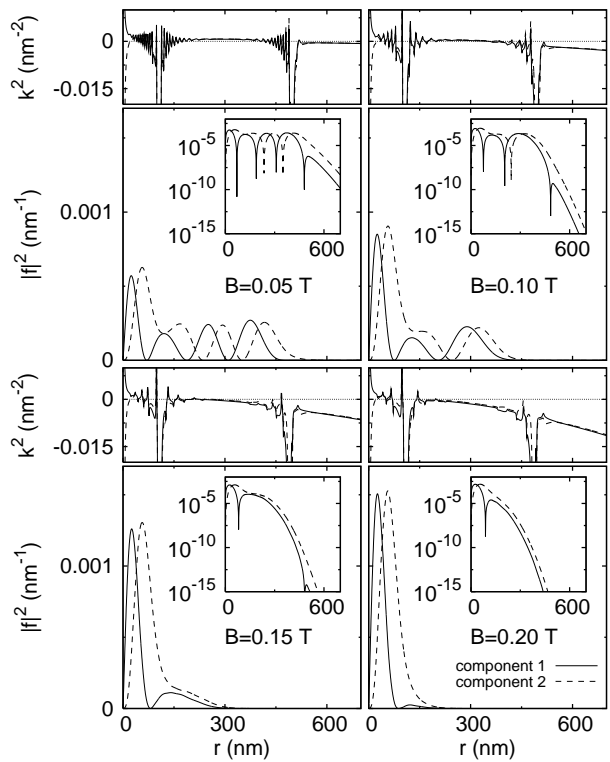}
\caption{Confinement-deconfinement transition. As Fig.~\ref{trans} but
for the potential shown in Fig.~\ref{potent}.}\label{realist}
\end{center}
\end{figure}
The ``confinement-deconfinement'' transition for the potential in
Fig.~\ref{potent} and a uniform magnetic field is shown in
Fig.~\ref{realist}. For all fields, the states have a peak near the centre
of the dot and an exponential tail. The transition occurs in the middle
distance region between these two features, $200 \alt r \alt 400$
nm.  For any potential, this region can be identified by computing
$k_i^2$. In Fig.~\ref{realist}, this varies rapidly because of the steps
in the potential but remains positive in the middle distance region when
B = 0.05 T. At this field, the oscillations in $|f_i|^2$ correspond to those
in Fig.~\ref{trans} but are much less rapid because
$V$, hence $k_i^2$ is smaller.
As the field increases the region of positive $k_i^2$ shrinks and a
transition to exponential behaviour occurs.  This is analogous to the
change of character seen in the Klein paradox for relativistic particles
with mass.

The occurrence of the transition is insensitive to the electrode geometry.
The one in Fig.~\ref{potent} has the advantage that the graphene is easy to
access but may be difficult to fabricate. However, similar transitions
occur in systems with disk or spherical central electrodes with the plate
and central electrode either above or below the graphene sheet. In
addition, calculations for model potentials with a well, barrier and flat
portion show that the occurrence of the transition is insensitive to the
model parameters. The only requirement is a region where $k_i^2$ changes
sign when $A_0$ increases. This is relatively easy to arrange.

$|f_i|^2$ in Fig.~\ref{realist} decreases by 3-4 orders of magnitude at
$r \approx 300$ nm when the character of the state changes from oscillatory
to exponential.  This large effect could be used to probe the transition
experimentally. For example, the decrease in $|f_i|^2$ causes a decrease
in the local density of states (LDOS) near the dot which could be detected
with scanning tunnelling microscopy.  The decrease in $|f_i|^2$ would also
cause a large decrease in the overlap of the dot state with states in
contacts at $r \approx 300$ nm. This could be detected by looking for a
change in transport through the dot state via diametrically opposite
contacts at $r \approx 300$ nm. Numerical calculations of the LDOS
for the potential in Fig.~\ref{potent} show that the state in
Fig.~\ref{realist} lies in a region of very low bulk density of states, 
with the dot state around 0.1-0.2 meV away from any other levels. This
suggests the dot state can be resolved experimentally in graphene samples
of sufficient quality.

In summary, confined and deconfined states occur in graphene dots in a
magnetic field. The states are confined when the scalar potential rises
slowly and deconfined when it rises rapidly. But when the scalar and vector
potentials have the same power law, the character of the states can be
controlled at will and a confinement-deconfinement transition occurs.  A
similar effect occurs in a realistic dot model. This is analogous to the
relativistic Klein paradox and may be experimentally observable.
The proposed system may be attractive because it allows graphene dots
to be formed in a uniform magnetic field.

It is a pleasure to acknowledge useful discussions with Professors H. Aoki,
H. Fukuyama and S. Tarucha. This work was supported by the UK Engineering
and Physical Sciences Research Council.


\begin{thebibliography}{99}
\bibitem{nanocrystal} The physics of nanocrystals is expected to be different.
\bibitem{reason1} Centred differences require
different numbers of grid points for $\chi_1$ and $\chi_2$ and lead to
unphysical solutions which are not smooth in the limit of zero step size.
\bibitem{reason2} The alternative choice, $\chi_1(R)=0$ is incompatible
with the discretization used in the present work.
\end{thebibliography}
\end{document}